\documentclass[prb,a4paper,twocolumn,floatfix]{revtex4-1}
\usepackage{color}
\usepackage[usenames,dvipsnames]{xcolor}
\usepackage{graphicx}
\usepackage{latexsym}
\usepackage[]{amsmath}
\usepackage{epstopdf}
\usepackage{amssymb}
\usepackage{array}

\usepackage{dcolumn}
\usepackage{bm}

\begin{document}

\title{On the temperature scaling behaviour of the linear magnetoresistance
observed in high-temperature superconductors}

\author{John Singleton$^{1,2}$}

\affiliation{$^1$National High Magnetic Field Laboratory, MS-E536, 
Los Alamos National Laboratory, Los Alamos, NM~87545, U.S.A.\\
$^2$University of Oxford, Department of Physics,
The Clarendon Laboratory, Parks Road, Oxford, OX1~3PU,
United Kingdom}

\begin{abstract}
An analytical model invoking
variations in the charge-carrier density is used to
generate magnetoresistance curves that are 
almost indistinguishable from
those produced by sophisticated numerical 
models. This demonstrates that,
though disorder is pivotal in causing linear magnetoresistance,
the form of the magnetoresistance thus 
generated  is insensitive to
details of the disorder.
Taken in conjunction with
the temperature ($T$) dependence of the zero-field
resistivity, realistic levels of disorder are
shown to be sufficient
to explain the linear magnetoresistance and 
field-$T$ resistance scaling observed in
high-temperature pnictide and cuprate superconductors.
Hence, though the $T$-linear zero-field resistance is a definite signature
of the ``strange metal'' state of high-temperature superconductors,
their linear magnetoresistance and its scaling is unlikely to be so.
\end{abstract}


\maketitle
\begin{figure}[t]
	\centering
	\includegraphics[width=5cm]{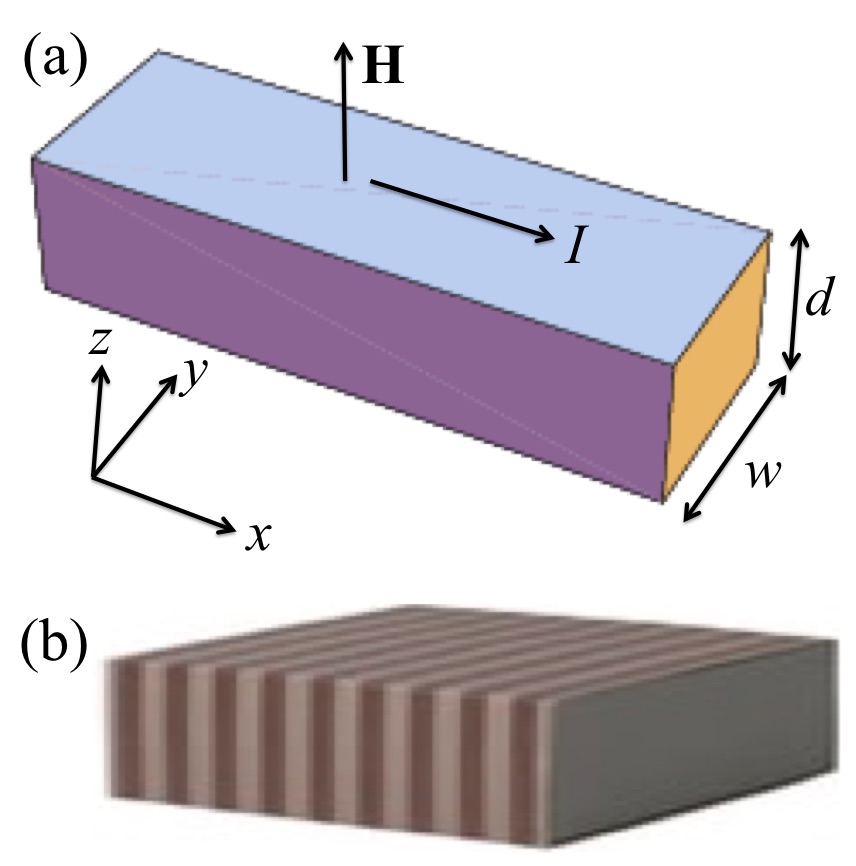}
	\sloppypar
	\vspace{-3mm}
	\caption{(a)~The geometry of the experiments modelled in this paper,
		carried out on bar-like samples.
		Arrows show the directions of the
		Cartesian $(x,y,z)$ axes; the point 
		$(x,y,z)=(0,0,0)$ is at the geometrical centre of the bar.
		The transverse magnetoresistance is obtained by measuring the voltage 
		between two contacts placed at the same height ($z$)
		on one of the faces (purple) parallel to the $xz$ plane.
		The current $I$ is directed along the bar $(x)$ axis, 
		perpendicular to the applied field ${\bf H} || z$.
		The bar is $w$ wide and $d$ thick.
		(b)~Cartoon of a bar-like sample consisting of 
		alternating layers that have $\kappa$ negative (brown)
		and positive (pink).}
	\vspace{-3mm}
	\label{fig1}
\end{figure}

In experimental {\it tours de force}~\cite{analytis1,analytis2,greg}, 
the magnetoresistance of pnictide and cuprate superconductors
- archetypal {\it strange metals} - has been measured in magnetic fields $\mu_0H$ of up to 92~T as a function of
temperature $T$. It was found that the
 transverse magnetoresistance $\rho$
[measured along the length of a bar-like sample with the current $I$
applied along the bar axis, 
$\perp {\bf H}$ - see Fig.~\ref{fig1}(a)]
exhibited interesting scaling behaviour with $H$ and $T$.
For instance, in Ref.~\onlinecite{analytis1}, when
$(\rho(H,T)-\rho_0)/T$, where $\rho_0$ is the residual resistivity, 
is plotted against $H/T$,
the data map onto a single curve, which tends to
a straight line ({\it i.e.} linear magnetoresistance) at larger values of $H/T$.
Any observation that gives a ``clue towards our... understanding of
the strange metal state in high-temperature 
superconductors~\cite{analytis1}'' is very welcome;
consequently, the linear magnetoresistance data have stimulated
theoretical models such as those in Refs.~\onlinecite{stringtheory,sachdev}.
For example, Ref.~\onlinecite{sachdev} invokes
a disordered strange metal consisting of itinerant electrons
interacting via random couplings with naturally formed ``quantum dots''
containing localized electrons; the $T$-linear zero-field
resistivity and some aspects of the magnetoresistance scaling
are reproduced~\cite{sachdev}.

However, linear magnetoresistance is far from being the
preserve of strange metals, having been
measured in many other systems for decades (see {\it e.g.}, 
\onlinecite{abrikosov1,abrikosov2,littlewood,stroud,probreferees}).
A particular issue was the 
observation of linear magnetoresistance in
``simple metals'' such as Al~[\onlinecite{bruls1,bruls2,bruls3}], 
whereas undergraduate text books~\cite{singleton} state that
this should {\it not} occur.
As reviewed in Ref.~\onlinecite{probreferees},
the necessary ingredient 
is disorder, the most sophisticated treatments
of which are
the {\it Random Resistor Network} (RRN)
model  and the {\it Effective Medium Theory} (EMT).
The former constructs a grid of four-terminal resistors, 
each with a varying random resistance; the latter posits a smoothly varying disorder potential that causes a continuous variation of the local conductivity. Both models belong to the same universality class and they are in reasonable agreement with data from a diverse range of materials~\cite{probreferees}.

Despite these successes, discussions at conferences in the 
past couple of years~\cite{conferences}
suggest that the RRN and EMT have little traction
in the strange-metal community. The reason is perhaps
that implementing these
models involves the specification
of inital parameters that are not straightforwardly related to the properties of real  materials, plus the application of randomness, followed by a full numerical calculation; fits to data are not trivial.

Therefore, the current paper has three purposes.
(i)~It describes a simple, analytical model for linear magnetoresistance due to disorder characterized by a single adjustable
parameter that is relatable to measurable material properties.
(ii)~It shows that the model reproduces the predictions of the more sophisticated RRN and EMT approaches to good accuracy,
thus permitting much simpler fits of theory to data.
(iii)~The good fit of the model to data enables the
$(H,T)$ scaling behaviour of the magnetoesistance 
of high-temperature superconductors
to be explained solely in terms of the expected disorder, 
without the need for explicitly 
``strange metal'' physics.

My starting point is the work of Bruls and
coworkers, who showed that variations in Hall voltage $V_{\rm H}$
along the length of a bar-like sample can result in linear 
magnetoresistance~\cite{bruls1,bruls2,bruls3}.
(For a simple metal with a spherical Fermi surface, populated
by a density $n$ electrons per unit volume, the Hall voltage
in a magnetic flux density $B$ is~\cite{singleton}
\begin{equation}
V_{\rm H} = -IB/(ned),
\label{Hally}
\end{equation} 
where $d$ is the sample thickness and 
$e$ is the magnitude of the charge of
the electron.)
Bruls {\it et al.} proved this idea by measuring
bars of very pure Al, machined so that $d$ 
(and hence $V_{\rm H}$ - see Eq.~\ref{Hally}) varied along their length~\cite{bruls1,bruls2,bruls3}.
An equivalent effect is obtained by varying $n$ along the
bar~\cite{bate}; I show here that this principle can
be applied to the cuprates and pnictides to generate 
linear magnetoresistance with the required scaling behaviour.

The first part of the derivation  
is similar to that in Ref.~\onlinecite{bate};
however, the system dealt with in Ref.~\onlinecite{bate}
is very different from the pnictides and cuprates; an 
exploration of linear magnetoresistance 
is not the primary purpose of Ref.~\onlinecite{bate};
and the relevant expression
is given in terms of parameters
that are not easily understood~\cite{whine}. 
Therefore I give a simplified, 
but complete form of the derivation (SI units,
contemporary notation) in the hope that
it will encourage others to play with the model and/or
develop calculations from a similar
starting point. 

I assume a bar-like sample fabricated from a simple metal
[Fig.~\ref{fig1}(a)]; 
the latter has a single, negatively charged carrier type of
number density $n$ and with effective mass $m^*$
(it is easy to show that the same result obtains for positively charged holes).
If {\bf H} is parallel to $z$ (as it is in most of the experiments~\cite{analytis1,greg})
it does not matter 
whether this material has a three-dimensional (spherical)
or two-dimensional (cylindrical) Fermi surface, as long as
the axis of the latter is also parallel to $z$.
The current runs along $x$, parallel to the bar long axis, and the field is applied
parallel to $z$. I assume that the sample
is not very magnetic, so that the magnetic flux density within it
and the applied field are roughly the same: {\it i.e.,} $B \approx \mu_0 H$.

The nature of the Lorentz force $(-e{\bf v}\times {\bf B})$ means that
there is no change to the carrier motion in the $z$ direction~\cite{singleton}, 
parallel to ${\bf B}$.
It is therefore sufficent to consider the following elements 
of the conductivity tensor ${\underline \sigma}$:
\[
\sigma_{xx} = \sigma_{yy} = \sigma^\prime = \frac{ne^2\tau}{m^*}\frac{1}{(1+\omega_{\rm c}^2\tau^2)}
\]
\vspace{-4mm}
\begin{equation}
\sigma_{yx} = -\sigma_{xy}=\sigma^\prime \omega_{\rm c}\tau.
\label{sigmas}
\end{equation}
Here, $\tau$ is the scattering time within the relaxation-time approximation and
$\omega_{\rm c} =\frac{eB}{m^*}$ is the cyclotron frequency~\cite{singleton}.

In the steady state, the continuity equation~\cite{singleton} 
is
\begin{equation}
\nabla . {\bf j} = 0,
\label{nothing}
\end{equation}
where ${\bf j} = {\underline \sigma} {\bf E}$ is the current density 
in the sample due to electric field ${\bf E}$.
Experiments~\cite{analytis1,analytis2,greg,bruls1,bruls2,bruls3} 
measure the voltages $V=V(x,y)$ at various points on the rod; 
remembering that ${\bf E} = -\nabla V$,
Eqs.~\ref{sigmas} and \ref{nothing} combine to yield
\[
\sigma^\prime \left[\frac{\partial^2V}{\partial x^2}+\frac{\partial^2 V}{\partial y^2}\right]+\frac{\partial V}{\partial x}
\left[\frac{\partial \sigma^\prime}{\partial x} +\frac{\partial (\omega_{\rm c}\tau\sigma^\prime)}{\partial y}\right]
\]
\vspace{-4mm}
\begin{equation}
+\frac{\partial V}{\partial y}
\left[\frac{\partial \sigma^\prime}{\partial y} -\frac{\partial (\omega_{\rm c}\tau\sigma^\prime)}{\partial x}\right]=0,
\label{bigass}
\end{equation}
where the possibility that 
$\sigma^\prime, ~~\omega_{\rm c}$ and/or $\tau$ might vary with $(x,y)$
has been introduced; the disorder and inhomogeneities
that might produce such variations are
a common feature of many models of linear magnetoresistance
in conventional metals and 
semiconductors~\cite{littlewood,stroud,probreferees,bruls1,bruls2,bruls3,bate}.

The pnictides~\cite{analytis1,analytis2} 
BaFe$_2$(As$_{1-x}$P$_x$)$_2$ and cuprates~\cite{greg} La$_{2-x}$Sr$_x$CuO$_4$
used in the linear magnetoresistance studies 
are alloys, and it is inevitable that some
inhomogeneity of composition (and hence $n$) will occur.
In addition, cuprates suffer from slight variations in oxygen stoichiometry,
again leading to variable $n$ (see {\it e.g.}, Ref.~\onlinecite{ramshaw}
and references therein). Finally,
the carrier density in both types of high-temperature
superconductor is much smaller than that
in conventional metals such as Al and Cu~\cite{analytis1,analytis2,greg,singleton,ramshaw}, 
potentially leading to much less effective screening
of disorder due to impurities and other defects~\cite{alloys}. 
Even in crystalline organic superconductors,
relatively clean systems compared to the cuprates,
this reduced screening
was shown to lead to an inhomogeneous carrier density~\cite{alloys}.
The current model therefore allows
$n$ to vary with position in the sample~\cite{bate};
other quantities such as $m^*$ and $\tau$ are 
assumed to be much less affected by inhomogeneities
and so are taken to be constants.

Eq.~\ref{bigass} could be attacked using a variety of
techniques. However, my motivation is
to derive a tractable expression for the 
magnetoresistance measured using
the voltage drop in the $x$ direction. 
I therefore consider a variation of
$n$ only in the $x$ direction
(the $\sim 10$~nm to $\sim 1~\mu$m
lengthscales
over which this happens will be discussed below). 
Separation of variables is then applied to
Eq.~\ref{bigass} to produce an analytical solution.
We hence choose~\cite{bate} 
\begin{equation}
n=n_0{\rm e}^{\kappa x}
\end{equation}
with $n_0$ and $\kappa \neq 0$ being constants.
Insertion of this into Equation~\ref{bigass} produces
\begin{equation}
\left[\frac{\partial^2 V}{\partial x^2}+\frac{\partial^2 V}{\partial y^2}\right]
+\kappa\left[\frac{\partial V}{\partial x}-\omega_{\rm c}\tau\frac{\partial V}{\partial y}\right] =0.
 \end{equation}
This has the solution 
$V=V_0{\rm e}^{-\kappa x}{\rm e}^{\kappa\omega_{\rm c}\tau y}+C$,
where $V_0$ and $C$ are constants; as 
Refs.~\onlinecite{analytis1,analytis2,greg} deal only
with voltage {\it differences} between contacts, 
$C$ can be set equal to $0$.
(Owing to the way in which the variation of 
$n$ with $x$ was chosen, 
this solution automatically satifies the 
boundary condition that no current should
flow out of the sample surfaces at 
$y=\pm \frac{w}{2}$ [Fig.~\ref{fig1}(a)].)
 
The fact that a current $I$ flows along the
sample bar (Figure~\ref{fig1}) is used to find $V_0$. We have
 \[
 I=d\int^{+\frac{w}{2}}_{-\frac{w}{2}}j_{x}{\rm d}y 
 \]
\vspace{-5mm}
\begin{equation}
 = \sigma^\prime\kappa V_0{\rm e}^{-\kappa x}(1+\omega_c^2\tau^2)d\int^{+\frac{w}{2}}_{-\frac{w}{2}}{\rm e}^{\kappa\omega_{\rm c}\tau y}{\rm d}y;
\end{equation}
some rearrangement gives $V_0$, yielding
\begin{equation}
V=\frac{I}{\sigma_0}\frac{1}{\kappa d w}\left[\frac{\frac{\kappa w \omega_{\rm c}\tau}{2}} {\sinh ( \frac{\kappa w \omega_{\rm c}\tau}{2})}\right]
{\rm e}^{-\kappa x}{\rm e}^{\kappa\omega_{\rm c}\tau y},
\label{fullmonty}
\end{equation}
where $\sigma_0 = n_0e^2 \tau/m^*$.
Note that the central bracket tends to 
the value 1 in a well-behaved fashion as $\omega_{\rm c} \rightarrow 0$.

\begin{figure}[t]
\centering
\includegraphics[width=9cm]{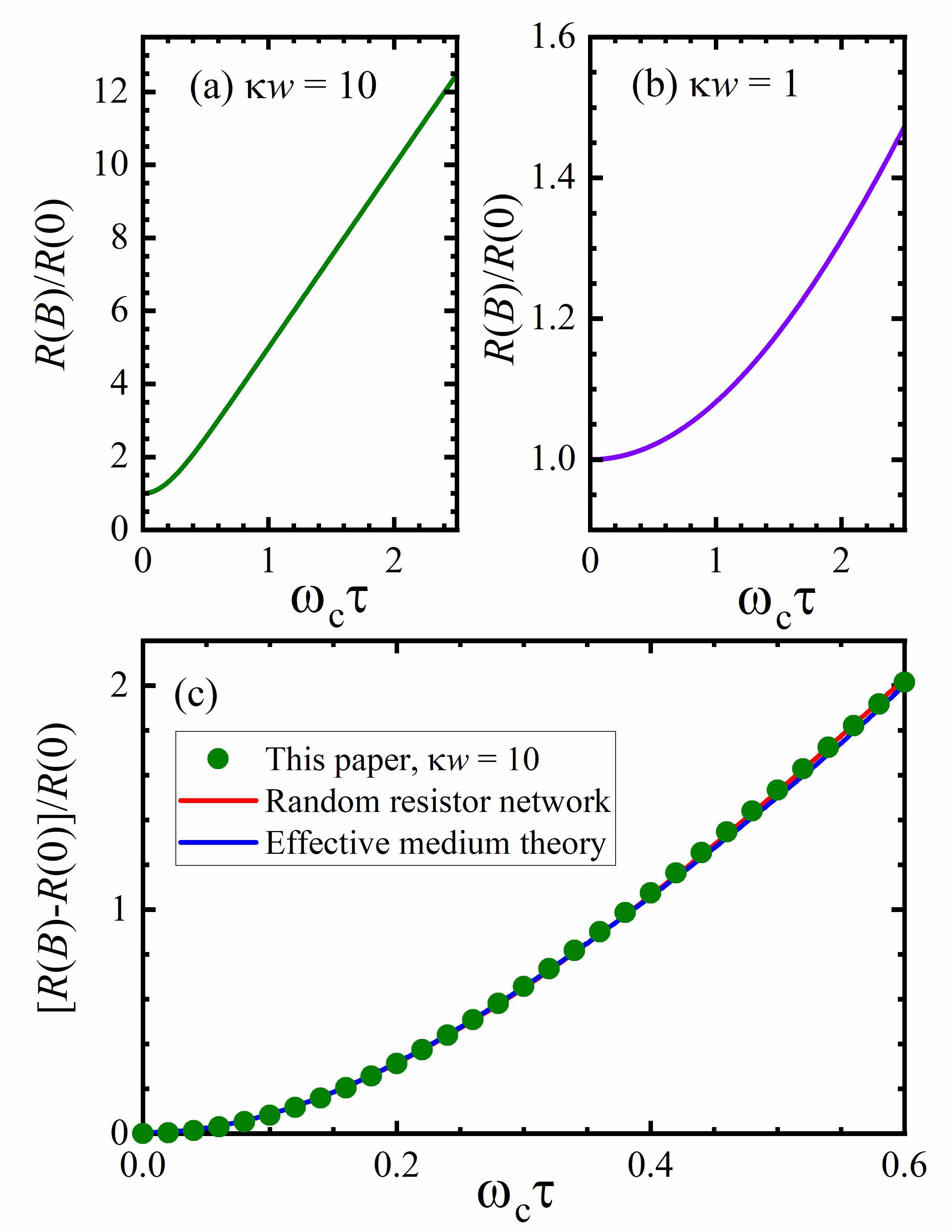}
\sloppypar
\vspace{-3mm}
\caption{(a,b)~Eq.~\ref{finalsolution}
plotted against $\omega_{\rm c}\tau$ 
for $|\kappa w| = 10$ [(a), green curve]
and $|\kappa w| = 1$ [(b), purple curve].
(c)~Comparison of part of the curve from (a) [green points]
with the predictions of the Random Resistor Network
model [RRN, red curve] 
and the  Effective Medium Theory [EMT, blue curve]~\cite{probreferees}.
Note that the EMT curve covers
the RRN curve for most of the figure.
Above $\omega_{\rm c}\tau\approx 0.3$,
the results of all three models lie
on straight lines [{\it i.e.}, all predict
linear magnetoresistance]
to good degree of accuracy.}
\vspace{-3mm}
\label{blah}
\end{figure}

In order to understand the 
data in Refs.~\onlinecite{analytis1,analytis2,greg},
consider voltages $V_1$ and $V_2$
at two contacts on
the same face [$\perp y$; Figure~\ref{fig1}(a)] of the bar at positions
$(x,y) = (x_1, \frac{w}{2})$ and $(x_2,\frac{w}{2})$
respectively.
Evaluating the voltage difference 
$\Delta V(B) = V_1-V_2$ at both $B=0$ and finite $B$,
and combining these two expressions gives
\vspace{-2mm}
\begin{equation}
\Delta V(B) = \Delta V(0)\frac{\kappa w \omega_{\rm c} \tau}
{1-{\rm e}^{-\kappa w \omega_{\rm c} \tau}}
\end{equation}
 or, in terms of the measured resistance, $R(B)= \Delta V(B)/I$,
\begin{equation}
 R(B) = R(0)\frac{\kappa w \omega_{\rm c} \tau}
{1-{\rm e}^{-\kappa w \omega_{\rm c} \tau}}, ~~~{\rm with}~~~\omega_{\rm c} =\frac{eB}{m^*}.
\label{oneside}
\end{equation}
For $\omega_{\rm c} \tau > 0$, the exponential
in the denominator of Eq.~\ref{oneside}
decreases with increasing $B$, leading
a dependence dominated by $\omega_{\rm c}\tau$
in the numerator; the magnetoresistance thus becomes
positive and linear. For $\omega_{\rm c} \tau <0$,
the exponental grows, leading to negative magnetoresistance.

The reason that Eq.~\ref{oneside} is asymmetric about
$B=0$ is that I have used $\kappa > 0$,
breaking the longitudinal symmetry of the bar.
However, unless the sample has been crafted with
an alloy composition that changes along the length of 
the bar~\cite{greg,bate},
one would expect $n(x)$ to vary both up and down
about some average value.
A minimalist approach is to imagine that
the rod consists of a stack of slabs along the $x$ direction 
in which $\kappa$
alternates between negative and positive
[Fig.~\ref{fig1}(b)].
For a large number of slabs
the resistance will be the average of
Equation~\ref{oneside} evaluated for $\kappa >0$ and $\kappa <0$:
\begin{equation}
R(B) = R(0)\frac{\kappa w \omega_{\rm c} \tau}{2}\left[
\frac{1}{1-{\rm e}^{-\kappa w \omega_{\rm c} \tau}}
-\frac{1}{1-{\rm e}^{\kappa w \omega_{\rm c} \tau}}
\right].
\label{finalsolution}
\end{equation}
This function, now symmetrical about $B=0$, 
is all that is needed for an understanding
of the data in Refs.~\onlinecite{analytis1,analytis2,greg}; 
it is plotted in Figs.~\ref{blah}(a,b).
Note that the magnetoresistance is a function of only 
the zero-field resistance and the product of two dimensionless
quantities, $\omega_{\rm c} \tau$ and $\kappa w$;
it will therefore follow {\it exactly the same temperature
dependence as $R(0)$ does}.
The shape is determined by the dimensionless quantities;
increasing $\kappa w$ moves the crossover from curved
to linear magnetoresistance
to lower values of $\omega_{\rm c} \tau$ [compare
Figs.~\ref{blah}(a) and (b)].

Fig.~\ref{blah}(c) compares the
lower part of the model curve shown in (a) 
[Eq.~\ref{finalsolution} with $|\kappa w| = 10$]
with numerical results 
from the RRN and EMT approaches. 
The RRN and EMT calculations employed the
parameters used to generate
the curves shown in Fig.~2  of Ref.~\onlinecite{probreferees};
the numerical scaling routine described in that paper~\cite{numcomment}
was used to match the $\omega_{\rm c}\tau$
range of the various models,
chosen to show the transition from
curved to linear magnetoresistance in detail.
There is very little difference between
the form of Eq.~\ref{finalsolution}
and the results of the 
more sophisticated numerical approaches~\cite{probreferees};
above $\omega_{\rm c} \tau \approx 0.3$,
the points from all three models lie
quite accurately on very similar straight lines.
In the rough and tumble of fitting experimental
data, such small differences will hardly matter.
In the prediction of linear magnetoresistance,
it seems that whilst disorder is of paramount
importance, the fine {\it details} of the disorder matter
little.\cite{fn1}

\begin{figure}[t]
	\centering
	\includegraphics[width=6.5cm]{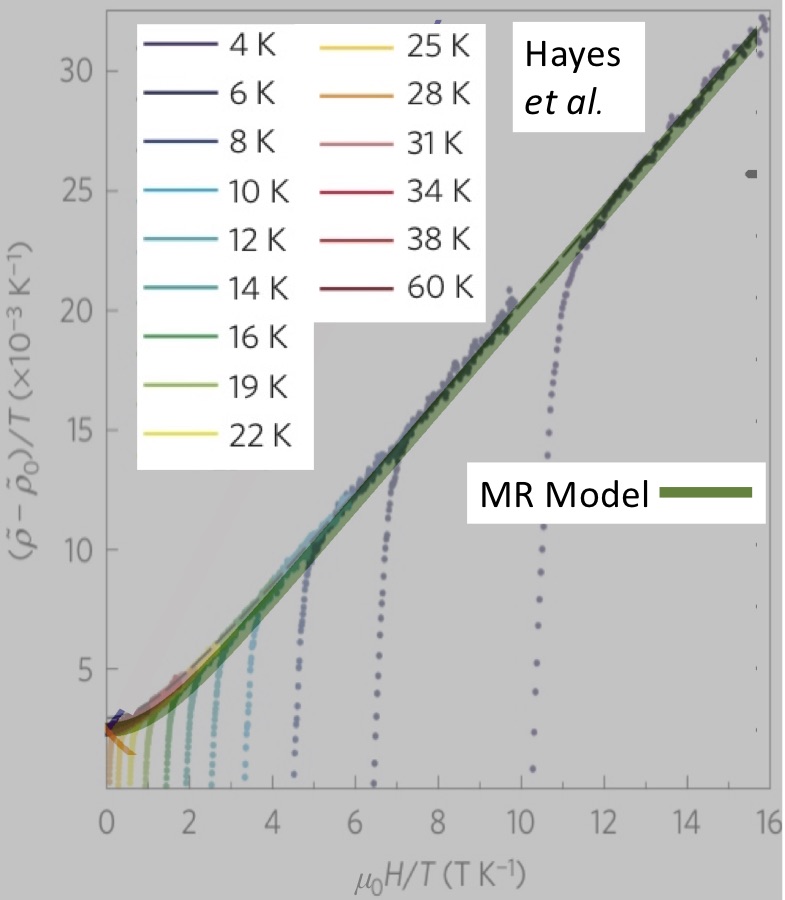}
	\sloppypar
	\vspace{-3mm}
	\caption{Figure from Ref.~\onlinecite{analytis1}
		showing magnetoresistance data 
		plotted as $(\rho -\rho_0)/T$ versus $\mu_0 H/T$;
		as discussed in the text, the latter quantity is
		proportional to $\omega_{\rm c} \tau$.
		The green curve is Eq.~\ref{finalsolution}
		employing (see text) $|\kappa w|=10$, with 
		$\omega_{\rm c} \tau \approx 2.5$ at
		$\mu_0H\approx 90$~T [the upper limit
		of the $x-$axis].}
	\vspace{-3mm}
	\label{blaty}
\end{figure}
The model curve in
Fig.~\ref{blah}(a) is already
very similar to data measured in Ref.~\onlinecite{analytis1}
[see Fig.~\ref{blaty}].
It therefore remains to explain the scaling behaviour.
Both pnictide and cuprate superconductors have
a normal-state, zero-field resistance that is 
written as~\cite{analytis1,analytis2,greg} 
\begin{equation}
\rho (0,T) = \rho_0 + AT
\end{equation}
where $\rho_0$ is a residual resistivity, assumed to
be unaffected~\cite{comment} by $H$ or $T$
in Refs.~\onlinecite{analytis1,analytis2}.
It is customary to subtract $\rho_0$ from the data
and treat the $H-$ and $T-$dependence of
what remains; I do the same.
For a fixed value of $\kappa w$,
Eq.~\ref{finalsolution}
takes the form $R(B,T)=R(0,T)f(\omega_{\rm c} \tau)$,
{\it i.e.} the zero-field resistance multiplied by a function
of only $\omega_{\rm c} \tau$.
If $R(0,T) = AT$, then 
\begin{equation}
R(B,T) =ATf(\omega_{\rm c} \tau).
\end{equation}
Therefore, the magnetoresistance scales linearly with $T$,
as observed~\cite{analytis1,analytis2,greg}. Moreover, if one 
plots $R(B,T)/T = Af(\omega_{\rm c} \tau)$,
the data should all collapse onto a single curve that
is a function of only $\omega_{\rm c} \tau$.

In the absence of temperature-dependent variations in carrier density
(not expected in a metallic system~\cite{singleton}),
the resistance will be proportional to a 
$T$-dependent scattering rate.
Put simply, 
$R(0,T) \propto 1/\tau$; since $R(0,T) = AT$, this implies that
$\tau \propto 1/T$.
Therefore, if $B$ is divided by $T$, one obtains a quantity
proportional to $\omega_{\rm c} \tau$. 
This is likely to be the reason
why the magnetoresistance data from Refs.~\onlinecite{analytis1,analytis2,greg}
divided by $T$ collapse onto a single curve when plotted 
versus $B/T$ [{\it e.g.}, Fig.~\ref{blaty}]. 

Having given a plausible explanation 
for the scaling behaviour in
Refs.~\onlinecite{analytis1,analytis2,greg},
it remains to discuss reasonable values for the
dimensionless quantities in the model
that can be used to fit such data.

\noindent
{\it Values of $\kappa w$:}~Based on general properties
of fairly disordered alloys~\cite{alloysbook}, one might expect
variations of $n$ of a few tenths of a percent to a few percent over the
lengthscale of typical microstructure,
which can range in size from $\sim 10$~nm to $\sim 1~\mu$m.
Therefore, $\kappa$ may be  $\sim 10^4-10^6~{\rm m}^{-1}$.
Typical samples used for pulsed-field transport measurements
tend to have a width $w\approx 50-500~\mu$m (author's observation).
Taking values somewhere in the middle of these ranges,
one obtains $\kappa w \sim 10$, as used in Fig.~\ref{blah}(a).

\noindent
{\it Values of $\omega_{\rm c} \tau$ in the cuprates and pnictides:}
Even good-quality cuprate superconductors that
exhibit Shubnikov-de Haas and de Haas-van Alphen
oscillations in high fields have relatively large scattering
rates; a typical sample of this sort~\cite{ramshaw} has 
$\omega_{\rm c} \tau \approx 1$ at 20~T.
The samples used in Refs.~\onlinecite{analytis1,analytis2,greg}
do not exhibit quantum oscillations;
they are chosen to be close to the quantum-critical point
and possess heavy masses and enhanced scattering.
I have guessed $\omega_{\rm c} \tau \approx 2.5$ at
$\mu_0H\approx 90$~T [the upper limit
of the $x-$axis of Figure~\ref{blaty}]. 
However, note that lower values of $\omega_{\rm c} \tau$
can be compensated for by higher values of $\kappa w$,
still well within the plausible limits discussed above.

Inserting these estimates of $\omega_{\rm c}\tau$
and $\kappa w$ into Eq.~\ref{finalsolution}
produces the green trace
overlaying the data in Fig.~\ref{blaty}. The model
matches the scaling
behaviour of the experiments well,
including the curvature seen at low $H$.

One further consequence of high scattering rates is that
the fine details of the Fermi-surface topology are
unlikely to be important in determining the transport properties;
hence a model that assumes a simple Fermi-surface toplogy,
such as the one presented here, may be adequate to
derive general qualitative features of the resistivity. 

Finally, I return to Ref.~\onlinecite{sachdev}, which models a diffusive
marginal Fermi liquid (MFL) to obtain a resistivity proportional to $T$.
It is notable that the MFL alone does not give linear 
magnetoresistance~\cite{sachdev};
only when a macroscopically disordered sample with 
domains of MFLs with varying densities of electrons
is introduced via EMT does the linear magnetoresistance
with the correct $(H,T)$ scaling
appear~\cite{sachdev}. 
This again supports the idea that the linear
magnetoresistance is not caused by
``strange metal'' physics; rather, it is merely a 
consequence of disorder.

In summary, an analytical model invoking realistic-sized 
variations in the charge-carrier density is used to
generate magnetoresistance curves that are 
almost indistinguishable from
those produced by the more sophisticated RRN and EMT
models. This demonstrates that,
though disorder is pivotal in causing linear magnetoresistance,
the form of the magnetoresistance thus 
generated is rather insensitive to the
microscopic details of the disorder~\cite{probreferees}.
Hence, the analytical expression derived [Eq.~\ref{finalsolution}]
can be used with confidence to 
fit the linear magnetoresistance and $(H,T)$
scaling reported in Refs.~\onlinecite{analytis1,analytis2,greg}.
The good correspondence between model and data, 
using plausible values for the two fit parameters,
shows that disorder of the size expected in alloys is
sufficent to explain the linear 
magnetoresistance and the way in which it
scales with the $T$-linear
zero-field resistivity that is a feature of pnictides and cuprates~\cite{analytis1,analytis2,greg}.
Therefore, though the latter $T$-linear zero-field resistance is
indeed a signature
of the ``strange metal'' state of high-temperature superconductors~\cite{sachdev},
the linear magnetoresistance {\it need not be}. 
\vspace{-4mm}

\section*{Acknowledgments}

This work was supported by
the US DoE Basic Energy Science Field Work Project {\it Science in 100 T},
and carried out at the National High Magnetic Field 
Laboratory, which is funded by NSF Cooperative Agreement 
DMR-1157490 and 1164477, the State of Florida and U.S. DoE. 
I thank the University of Oxford 
for the provision of a Visiting Professorship that permitted
calculations that underpin this paper. 
Paul Goddard, Stephen Blundell, Andrea Schmidt,
Arkady Shekhter, Nikola Maksimovich 
and Neil Harrison are thanked for valuable discussions.I am grateful
to Ross McDonald for allowing me to reproduce the
results from Ref.~\onlinecite{analytis1} plotted in Fig.~\ref{blaty}.


\begin{thebibliography}{99}
\bibitem{analytis1}
Ian M. Hayes, Ross D. McDonald, Nicholas P. Breznay, Toni Helm, 
Philip J. W. Moll, Mark Wartenbe, Arkady Shekhter and James G. Analytis,
{\it Nature Physics}, {\bf 12}, 916 (2016).
\bibitem{analytis2}
Ian M. Hayes, Zeyu Hao, Nikola Maksimovic, Sylvia K. Lewin, Mun K. Chan, Ross D. McDonald, B. J. Ramshaw, Joel E. Moore, James G. Analytis,
Phys. Rev. Lett. {\bf 121}, 197002 (2018).
\bibitem{greg}
P. Giraldo-Gallo, J. A. Galvis, Z. Stegen, K. A. Modic, F. F. Balakirev, J. B. Betts, X. Lian, 
C. Moir, S. C. Riggs, J. Wu, A. T. Bollinger, X. He, I. Bozovic, B. J. Ramshaw, R. D. McDonald, G. S. Boebinger, A. Shekhter,
{\it Science}, {\bf 361} 479 (2018).
\bibitem{stringtheory}
S. Cremonini, A. Hoover, and L. Li, {\it J. High Energ. Phys.} {\bf 2017}, 133 (2017). 
\bibitem{sachdev}
A.A. Patel, J. McGreevy, D.P. Arovas and S.~Sachdev,
Phys. Rev. X {\bf 8}, 021049 (2018).
\bibitem{abrikosov1}
A.A.~Abrikosov, {\it Europhysics Letters}, {\bf 49}, 789 (200).
\bibitem{abrikosov2}
A.A. Abrikosov, {\it J. Phys. A: Math. Gen.} {\bf 36} 9119 (2003).
\bibitem{littlewood}
M.M. Parish and P.B. Littlewood, {\it Nature} {\bf 426},
162 (2003).
\bibitem{stroud}
D. Stroud, {\it Phys. Rev. B} {\bf 12}, 3368 (1975).
\bibitem{probreferees}
Navneeth Ramakrishnan, Ying Tong Lai, Silvia Lara, Meera M. Parish, and Shaffique Adam, Phys. Rev. B {\bf 96}, 224203 (2017) 
\bibitem{bruls1}
G.J.C.L. Bruls, J. Bass, A.P. van Gelder, H. van Kempen and P. Wyder,
Phys. Rev. Lett. {\bf 46}, 553 (1981).
\bibitem{bruls2}
G.J.C.L. Bruls, J. Bass, A.P. van Gelder, H. van Kempen and P. Wyder,
{\it Phys. Rev.} {\bf B 32}, 1927 (1985)
\bibitem{bruls3}
G.J.C.L. Bruls, Ph. D. Thesis, Katholieke Universiteit Nijmegen
(Krips Repro, Meppel, 1985).
\bibitem{singleton}
See {\it e.g.,} Neil W. Ashcroft  and N. David Mermin,
{\it Solid State Physics} (Holt, Rinehart and Winston, 1976)
or J. Singleton
{\it Band theory and electronic properties of solids}
(Oxford University Press, Oxford, 2002), Chs. 1 and 10.
\bibitem{conferences}
For example, satellites of the International Conference on Strongly Correlated Electron Systems 2019 (SCES '19)
and the American Physical Society March Meeting 2019.
\bibitem{bate}
R.T. Bate and A.C. Beer, {\it J. Appl. Phys.} {\bf 32}, 800 (1961).
\bibitem{whine}
So much so that the current author was forced to rederive
all of the expressions to check that he understood the
various ill-defined $\rho$ parameters.
\bibitem{ramshaw}
B. J. Ramshaw, Baptiste Vignolle, James Day, Ruixing Liang, W. N. Hardy, 
Cyril Proust and D. A. Bonn,
{\it Nature Physics}, {\bf 7}, 234 (2011).
\bibitem{alloys}
 J. Singleton, N. Harrison, C.H. Mielke, 
J.A. Schlueter and A.M. Kini,
{\it J. Phys.: Condens. Matter}, {\bf 13}, L899 (2001).
\bibitem{numcomment}
In Fig.~2 of Ref.~\onlinecite{probreferees},
magnetoresistance curves with
disorder parameters
$\eta =0.1$ and $\eta =10$ from both the RRN
and EMT models 
are indistinguishable to the naked eye,
so $\eta = 1$ is used here.
The normalization routine of
Ref.~\onlinecite{probreferees} looks 
for the point at which the low-$B$
part of the magnetoresistance deviates from a parabola
and uses this for field-axis scaling.
\bibitem{fn1}
Similarly, in a 
real sample, the composition
variations will be more random than the simple case
described in this paper.
One could deal with this
by evaluating Eq.~\ref{finalsolution} 
for many different values of $\kappa$ and
averaging the results.
However, a few quick calculations (left as an exercise for the reader)
will show that this does very little to
change the form of the magnetoresistance.
\bibitem{comment}
The fact that the resistivity is treated
in Refs.~\onlinecite{analytis1,analytis2,greg} as
the sum of two completely independent
terms suggests that samples look like a 
series-connected network of ``dead'' regions
(giving the residual resistivity, $\rho_0$),
and ``alive'' regions that possess quasiparticles
that respond to field and temperature.
This is part of the reasoning behind treatments
such as that of Ref.~\onlinecite{sachdev}.
\bibitem{alloysbook}
See {\it e.g.}, {\it Fundamentals of Metallurgy}
(1st Edition), Ed. S Seetharaman
(Woodhead Publishing, Cambridge, 2005).
\bibitem{contacts}
In some cases, experiments are carried out
using contacts made using wires connected to a smear of
silver paint or silver epoxy across the
whole width of the top surface ($\perp z$- see Fig.~\ref{fig1})
of a bar (author's observation).
Such contacts frequently have a resistance $\sim 10-100$
times the four-contact resistance of the
bar and hence do not represent an equipotential
region. Instead, they will detect an average of the
voltages from different $y$ values
across the bar, meaning that the form of the 
magnetoresistance will be essentially the same as Eq.~\ref{finalsolution}.
Note that, because the current model considers only 
one-dimensional variations in $n$,
it cannot be used to predict quantitative
differences between different contact geometries
({\it e.g.}, computing the relative magnitudes of voltages from
pairs of contacts on the bar faces perpendicular to $z$ and to $y$). 
\end{thebibliography}
\end{document}